**FMR studies of exchange-coupled multiferroic polycrystalline $Pt/BiFeO_3/Ni_{81}Fe_{19}/Pt$ heterostructures**

**Authors**

Jamal Ben Youssef, Jérôme Richy, Nathan Beaulieu, Tony Hauguel , David T. Dekadjevi , Jean-Philippe Jay, David Spenato and Souren P. Pogossian[1]

Laboratoire de Magnétisme de Bretagne, UBO/UEB/CNRS, EA 4522, 29285 Brest

[1]corresponding author email: pogossia@univ-brest.fr

## Abstract

An experimental study of the in-plane azimuthal behaviour and frequency dependence of the ferromagnetic resonance field and the resonance linewidth as a function of $BiFeO_3$ thickness is carried out in a polycrystalline exchange-biased $BiFeO_3/Ni_{81}Fe_{19}$ system. The magnetization decrease of the $Pt/BiFeO_3/Ni_{81}Fe_{19}/Pt$ heterostructures with $BiFeO_3$ thickness deduced from static measurements has been confirmed by dynamic investigations. Ferromagnetic resonance measurements have shown lower gyromagnetic ratio in a perpendicular geometry compared with that of a parallel geometry.

The monotonous decrease of gyromagnetic ratio in a perpendicular geometry as a function of the $BiFeO_3$ film thickness seems to be related to the spin-orbit interactions due to the neighbouring Pt film at its interface with $Ni_{81}Fe_{19}$ film.

The in-plane azimuthal shape of the total linewidth of the uniform mode shows isotropic behaviour that increases with $BiFeO_3$ thickness. The study of the frequency dependence of the resonance linewidth in a broad band of 3–35 GHz has allowed the determination of intrinsic and extrinsic contributions to the relaxation as function of $BiFeO_3$ thickness in perpendicular geometries. In our system the magnetic relaxation is dominated by the spin-pumping mechanism due to the presence of Pt. The insertion of $BiFeO_3$ between Pt and $Ni_{81}Fe_{19}$ attenuates the spin-pumping damping at one interface.

## Introduction

In recent information storage devices, multiferroics [1-4] are among the most attractive materials uniting two or more ferroic properties in a single phase. They have high technological impact in a great number of applications, including electrically-controlled microwave phase shifters, filters, resonators, magnetically-controlled electro-optic or piezoelectric devices, broad band magnetic field sensors etc [5-10]. Multiferroic bismuth iron oxide $BiFeO_3$ (BFO) is ferroelectric and antiferromagnetic (AF) at room temperature. Exchange-coupled BFO/ferromagnetic bilayers have attracted a lot of attention in the last decade due to the strong dependence of the magnetic spin configuration of these systems on microstructural properties. It is well known that local uncompensated spins at the interface are responsible for the appearance of exchange bias [11]. Therefore, the ferromagnetic interfacial exchange coupling with a BFO film is expected to depend on interfacial and inner microstructural characteristics induced by the growth technique and growth parameters used. The growth process may be used to make significant changes in BFO and FM since it may affect the exchange coupling. While the exchange coupling between bulk or epitaxial AF BFO and a ferromagnetic film at room temperature was reported by several authors [12-15], there has been little work concerning the exchange bias in polycrystalline BFO/ferromagnetic heterostructures [16-18]. In a recent paper we presented a systematic experimental study of the azimuthal evolution of static magnetic properties of exchange-coupled polycrystalline Pt/BFO/$Ni_{81}Fe_{19}$/Pt heterostructures with BFO thickness, in which we observed the presence of biquadratic, uniaxial and unidirectional anisotropies with a misaligned anisotropy axis [18].

Ferromagnetic resonance (FMR) is a reliable spectroscopic technique for probing the magnetic anisotropy of ferromagnetic films since it measures directly the internal magnetic fields in the system. As a result, it produces complete characterization of the magnetic anisotropy. Moreover, it allows an exact and direct determination of the saturation magnetization, as well as the gyromagnetic ratio, which in turn gives information on spin-orbit coupling and orbital to spin magnetic moment ratio ($\mu_L/\mu_S$) [19-20]. For this reason we performed FMR measurements to improve our understanding of the physical mechanisms of exchange coupling and complex magnetic anisotropy in BFO/$Ni_{81}Fe_{19}$ heterostructures as a function of BFO thickness.

It should also be noted that despite a large number of studies of dynamic behaviour of many exchange-biased systems, only a few concern BFO/ferromagnetic systems [21-25]. Crane et al. [22] studied the magnetic anisotropy of nanostructures composed of ferrimagnetic $NiFe_2O_4$ pillars in a multiferroic $BiFeO_3$ matrix using the ferromagnetic resonance technique. They showed that the uniaxial magnetic anisotropy in the growth direction changes sign as film thickness is increased. Magaraggia et al. [23] studied in-plane and out-of-plane anisotropies in multilayer $La_{1-x}Sr_xMnO_3$/BFO by FMR; however, they found no evidence of exchange bias in these systems. Guo et al. [24] used FMR to probe the magnetic anisotropy of a self-assembled system composed of $CoFe_2O_4$ nanopillars within a BFO matrix. Their results suggest the presence of four-fold symmetry of FMR spectra that they associate with the four-fold symmetry of $CoFe_2O_4$ nano-crystals.

In our present study, we make a detailed examination of the in-plane and out-of-plane FMR response of exchange-coupled Pt/BFO/Py/Pt nanometric structures, where Py (permalloy) denotes $Ni_{81}Fe_{19}$. We draw particular attention to the azimuthal behaviour and the frequency dependence of the FMR field and its linewidth in a 3–35 GHz broad band, as a function of BFO thickness in a polycrystalline exchange-biased Pt/BFO/$Ni_{81}Fe_{19}$/Pt system. These structures can also be valuable for magnonic studies since the FMR technique allows injection of spin current into the Pt buffer and Pt cap layers via magnetization precession in the $Ni_{81}Fe_{19}$ film [26].

This paper is organized as follows. In the first section we present briefly the sample growth, and characterization of nonmagnetic properties such as morphology and microstructure. In section 2, we resume concisely the static magnetic measurement results and typical physical trends. In section 3, the anisotropy properties of our samples are examined based on FMR field measurements. Section 4 treats the analysis of magnetic relaxation measured in in-plane and perpendicular geometries, followed by the conclusion.

## 1. Sample preparation and characterization

The Si/Pt(14nm)/BFO($t_{BFO}$)/Py (10nm)/Pt(10nm) heterostructures, investigated with BFO thickness $t_{BFO}$ ranging from 0 nm to 85 nm, were grown on Si(100) substrate by conventional Radio Frequency sputtering (Leybold Z550), under a static magnetic field $H_{dep}$ = 300 Oe applied in the plane of the film and at a base pressure of $10^{-7}$ mbar. The deposition field direction will be considered as the reference direction. Samples were annealed for one hour at 500°C under air without any applied magnetic annealing field. This deposition procedure allowed us to obtain exchange-coupled polycrystalline BFO/Py heterostructures as demonstrated in our previous work [16]. The investigated BFO thicknesses were $t_{BFO}$ = 0, 29, 39 and 85 nm. As a reference, we deposited a Py(10 nm)/$Al_2O_3$(10 nm) thin film structure without Pt and BFO. The capping film of $Al_2O_3$ was used to prevent oxidation.

The longitudinal magnetization parallel to the field direction was probed by vibrating sample magnetometry (VSM). A shifted hysteresis loop, at room temperature, was observed for $t_{BFO}$ = 29, 39 and 85 nm [Figs. 1 (a)]. Structural properties were probed by X-Ray diffraction and the dependence of morphology on BFO thickness analysed using Atomic Force Microscopy (AFM). The relative intensity of the Bragg peaks was in good agreement with the BFO powder pattern, indicating a lack of preferential orientation.

For thin BFO layers, a smooth surface was observed, with a root mean square roughness (RMS) of about 1 nm, while for thicker BFO films (>29 nm), a steep increase of RMS was observed that saturates at a value of about 15 nm, as shown in previous studies [16,18].

## 2. BFO thickness dependence of the hysteresis loop

The magnetization reversal along in-plane parallel ($||H_{dep}$) and in-plane perpendicular ($\perp H_{dep}$) to the deposition magnetic field ($H_{dep}$) direction are shown on Fig. 1(a) as measured for different BFO thicknesses. In such heterostructures, the hysteresis loop shift defines the exchange bias field $H_{eb}$ [Figs. 1(a)][16,18].

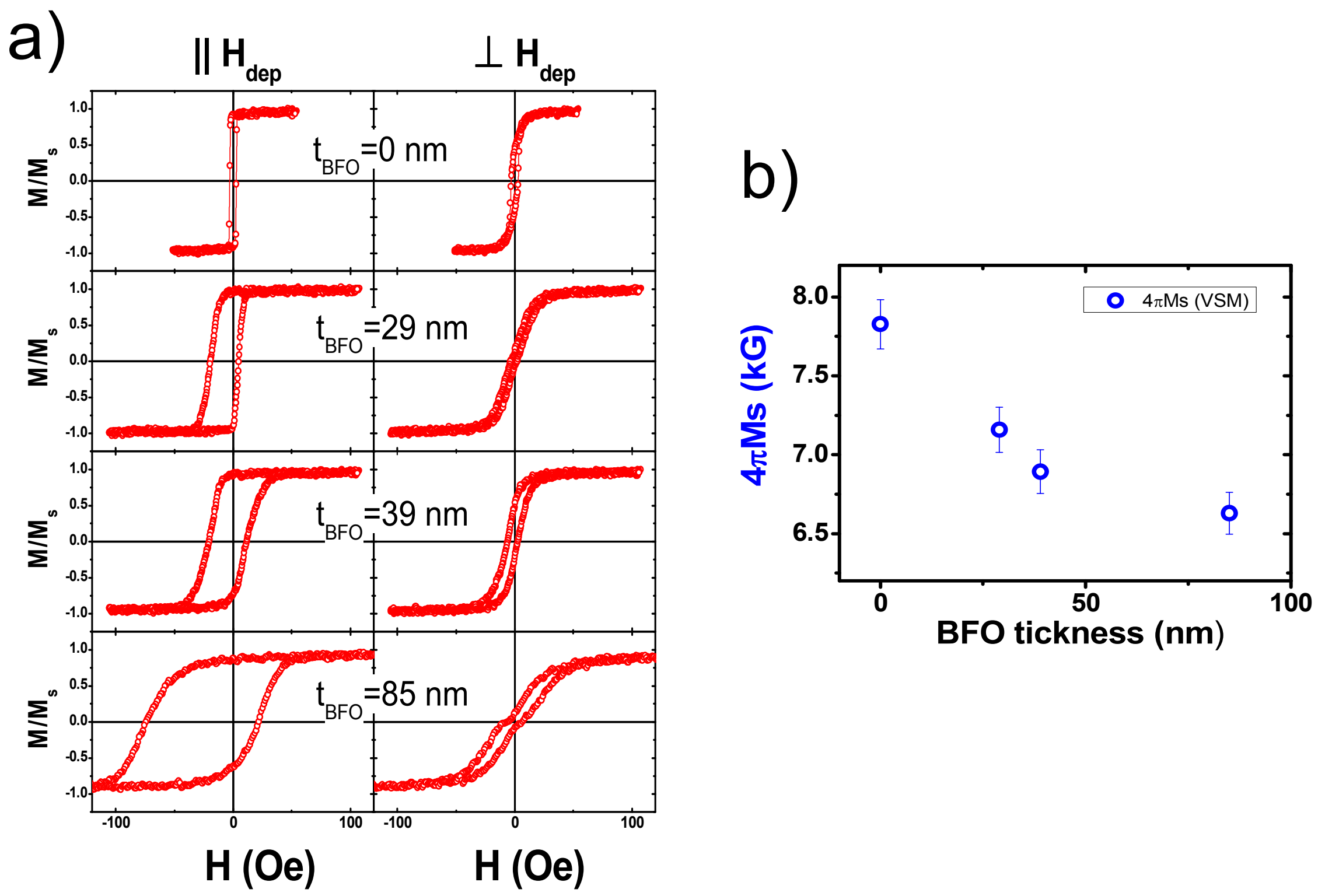

**Fig. 1** : (a) *Hysteresis loop along the direction parallel (left column) and in-plane perpendicular to the deposition field axis (right column) for BFO thicknesses* $t_{BFO}$ *= 0, 29, 39 and 85 nm. (b) Behaviour of saturation magnetization* $4\pi M_s$ *from static VSM measurements as a function of BFO thickness.*

As expected for the sample without BFO [*Pt(14nm)/Py(10nm)/Pt(10nm)*], the magnetization hysteresis curve does not show any loop shift. The coercive field of this sample is small (~ 6 Oe). The exchange bias appears for a critical thickness of BFO lying below 29 nm, as shown in previous studies [16,18,27].

For samples with $t_{BFO} \geq 29$ nm, we observe a typical magnetic behaviour of exchange-coupled systems, i.e., the presence of a hysteresis loop shift and a coercive field enhancement in the deposition-field direction. The observed magnetization two-step reversal in the in-plane direction perpendicular to the $H_{dep}$ axis for $t_{BFO}$ = 85 nm can be attributed to the biquadratic anisotropy [Fig. 1(a)]. For other thicknesses, the two-step reversal is not well pronounced, but that does not mean that biquadratic anisotropy is absent from these samples. According to Tang et al. [28], the exchange-bias effect described by unidirectional anisotropy is also accompanied by induced uniaxial and biquadratic in-plane anisotropy contributions in polycrystalline Fe/MnPd exchange-biased bilayers. We have already demonstrated in our previous work that in polycrystalline Pt/BFO/Py/Pt heterostructures grown on a Pt buffer layer there are three types of contributions to the magnetic anisotropy: unidirectional, uniaxial and biquadratic anisotropies [16,18]. The balance of these three anisotropies varies as a function of BFO thickness. Consequently, the shape of hysteresis curves may change considerably as a function of ratios of unidirectional, uniaxial, and biquadratic anisotropy fields. The presence of uniaxial and biquadratic anisotropies may be due to the complex spin configuration and local random coupling processes at the atomically rough AF/ferromagnetic interface [28-31].

Measured saturation magnetization [Fig. 1(b)] for the reference sample Pt/Py(10 nm)/Pt is $4\pi M_s$=7.8 kG, which is much smaller than that of the bulk Py ($4\pi M_s$=10 kG). As shown in Fig. 1(b), the saturation magnetization ($4\pi M_s$) measured by VSM decreases as a function of BFO thickness. A similar behaviour was observed in other systems such as a Py/Ta system [32], Ta/Py bilayers [33] and $Al_2O_3$/Py/$Al_2O_3$ trilayers [34]. In reference [32], the steep decrease of the saturation magnetization below 20 nm of Py thickness was attributed to the existence of an inhomogeneous interfacial layer at different interfaces. Ounadjela et al. [33], have associated the strong dependence of $4\pi M_s$ on the thickness of Ta/Py bilayers with the existence of inhomogeneous interfacial layer at the film surface Ta/Py and Py/substrate interfaces. In $Al_2O_3$/$Ni_{80}Fe_{20}$/$Al_2O_3$ trilayers [34], the reduction of magnetization was attributed to the change of stoichiometry due to the selective oxidation of Fe at the Py film interfaces. The decrease of the saturation magnetization ($4\pi M_s$) of our samples as a function of BFO thickness (Fig. 1(b)) may be associated with the evolution of the interfacial parameters such as interdiffusion or RMS roughness [18]. In order to investigate in detail the magnetization behaviour as a function of BFO thickness we have carried out dynamic studies using the ferromagnetic resonance technique.

## 3. Anisotropy study

The FMR field ($H_{res}$) and linewidth ($\Delta H$) are sensitive to the anisotropy of the film, including surface anisotropy, magnetic interactions between different elements of the magnetic heterostructure and interfaces interactions such as exchange bias, and magnetic and structural inhomogeneities in the sample. In the present section, in order to study the magnetic anisotropy of polycrystalline exchange-coupled Pt/BFO/Py/Pt heterostructures, FMR field and resonance linewidth measurements were carried out at 6 GHz frequency using a highly sensitive broadband resonance spectrometer with a nonresonant microstrip transmission line. The FMR is measured via the derivative of microwave power absorption using a small Radio Frequency (RF) exciting magnetic field. Resonance spectra were recorded with in-plane (Fig. 2(a)) and out of plane (Fig. 2(b)) oriented static magnetic field relative to the deposition field direction. The static magnetic field was swept from its highest initial positive value towards negative values [35]. Our experimental device allows the determination of both positive ($H_+$) and negative ($H_-$) resonance fields. The amplitude of modulation of the RF field $h_{rf}$ is appreciably smaller than the FMR linewidth : $h_{rf} << \Delta H$. The amplitude of RF exciting field $h_{rf}$ is evaluated to be about $h_{rf} \approx 2$ mOe, which corresponds to the linear response regime. A phase-sensitive detector with lock-in detection was used. The field derivative of the absorbed power, *dP/dH,* where $P = \frac{1}{2}\omega\chi'' h_{rf}^2$ ($\chi''$ being the imaginary part of the susceptibility of the uniform mode, $\omega=2\pi f$, and $f$ - the frequency of the exciting field) is proportional to the field derivative of the imaginary part of the susceptibility $d\chi''/dH$. The resonance field corresponds to the zero crossing of $d\chi''/dH$. The FMR peak to peak linewidth $\Delta H$ is given by the field interval between the extrema of $d\chi''/dH$ [35].

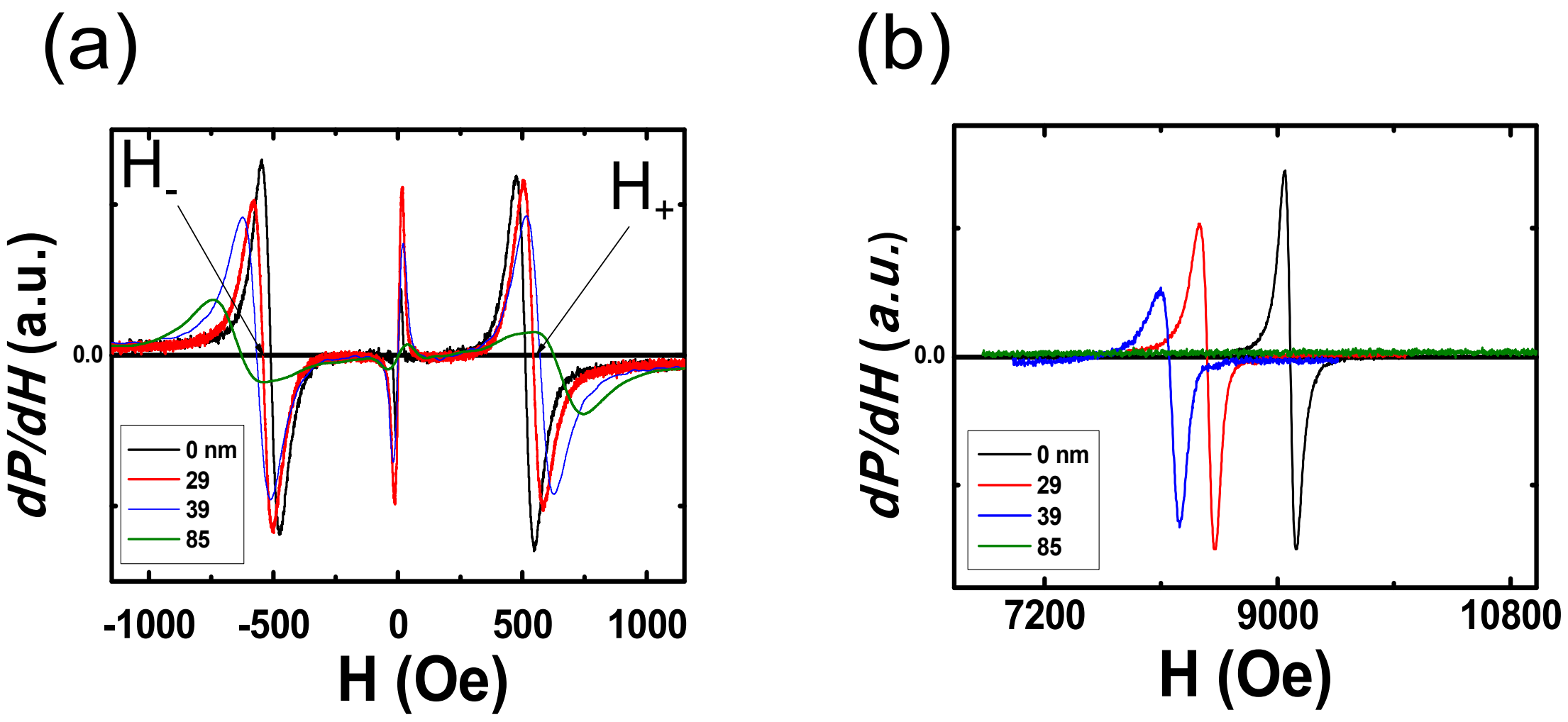

**Fig. 2** : *FMR power absorption field derivative dP/dH versus magnetic field H for a Pt/BFO/Py/Pt system for in-plane (a) and out- of-plane perpendicular (b) geometries.*

The considered exchange coupled samples with thicknesses $t_{\mathrm{BFO}}$ = 0, 29, 39 and 85 nm display a series of power absorption field derivative curves (in arbitrary units) for both in-plane and out-of-plane perpendicular geometries [Figs. 2(a) and 2(b)] obtained at room temperature. In Fig. 2(a), a displacement of the FMR field and an enhancement of the resonance linewidth with the increase of BFO thickness along the deposition field axis are observed in in-plane geometry. A strong decrease is observed in the intensity of the field derivative of the imaginary part of the susceptibility for $t_{\mathrm{BFO}}$=85 nm.

Apart from the two main uniform mode peaks corresponding to the two polarities ($H_+$, $H_-$) of applied magnetic field sweeps, there is a low field ($H$~0) absorption peak for all thicknesses. These low-field dynamic responses are different from the usual FMR uniform mode in the saturated magnetization state, which have a 180° phase shift with respect to the FMR signal of uniform mode [36]. They are attributed to low field absorption in the unsaturated magnetization state of the thin films owing to magnetic inhomogeneities associated with the domain structure [36-37]. The high field FMR resonance occurs for magnetic fields values much higher than the saturating field. The resonance field decreases when $t_{\mathrm{BFO}}$ also decreases.

The FMR absorption spectra in perpendicular geometry are shown in Fig. 2(b). In this figure, a decrease of the FMR field and enhancement of the resonance linewidth ($\Delta H$) is observed with increasing BFO thickness. No signal has been observed for $t_{\mathrm{BFO}}$ = 85 nm [Fig. 2(b)].

The gyromagnetic ratio $\gamma$ and the effective magnetic field $H_{eff}$ can be determined from frequency-dependent FMR measurements in out-of-plane perpendicular geometry. In this geometry, there is a linear relationship between $H_{res}^{\perp}$ and microwave frequency $f = \omega / 2\pi$ [38]

$$\omega = \gamma\left(H_{res}^{\perp} + H_{eff}^{\perp}\right), \tag{1}$$

The effective field $H_{eff}^{\perp} = -4\pi M_s + \dfrac{4K_{\perp}}{M_s t_F}$ is related to the surface anisotropy $K_{\perp}$ and to the saturation magnetization $4\pi M_s$. The thickness of Py film is noted by $t_F$.

When the anisotropy fields are negligible with respect to $4\pi M_s$ and to the ferromagnetic resonance field, the in-plane resonance field $H_{res}^{||}$ is found from [38]

$$\left(\frac{\omega}{\gamma}\right)^2 = H_{res}^{||}\left(H_{res}^{||} + 4\pi M_s\right). \qquad (2)$$

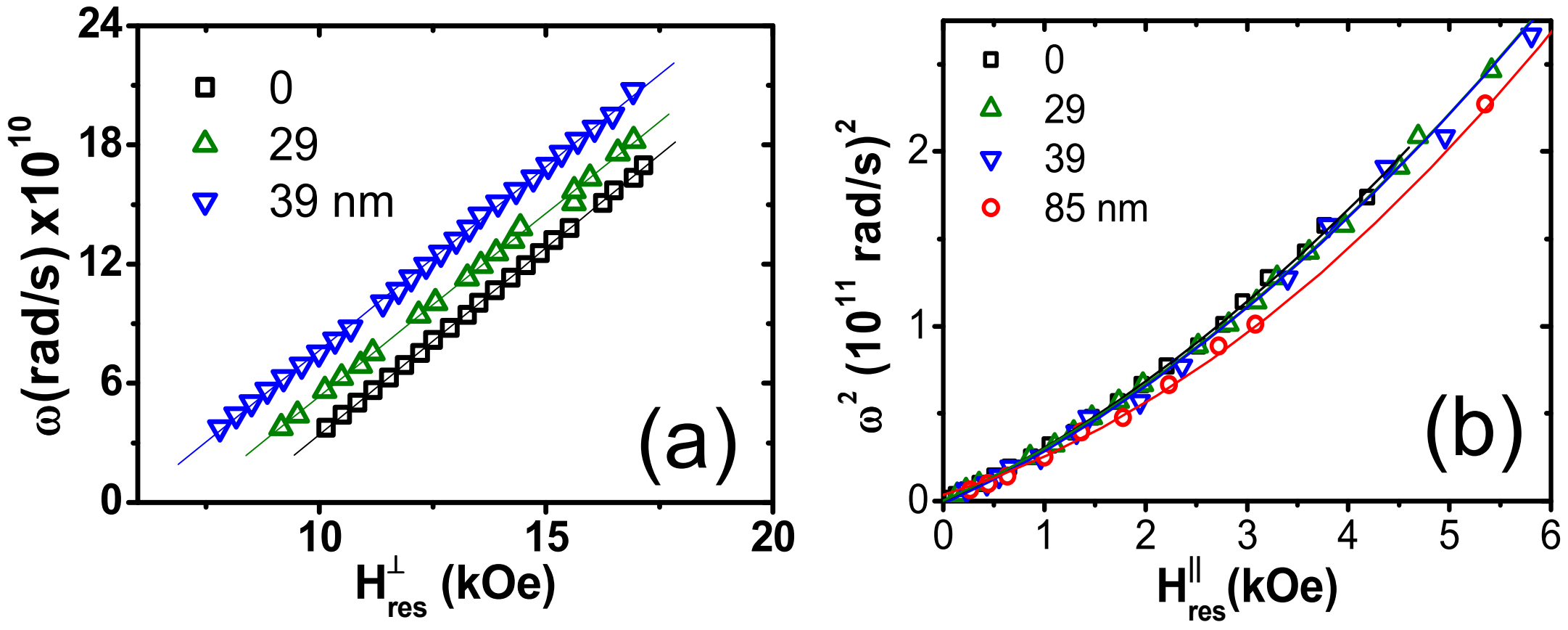

**Fig. 3** : *Evolution of the FMR field in out-of-plane perpendicular [(a): $H_{res}^{\perp}$] and in plane planar [(b): $H_{res}^{||}$] geometries as a function of resonance frequency for different BFO thicknesses*

Fig. 3(a) confirms the linear dependence of $H_{res}^{\perp}$ on the microwave frequency *f*. From that linear dependence, and from the Eq. (1), the gyromagnetic ratio $\gamma$ and the effective field $H_{eff}^{\perp}$ were determined. The effective field, $H_{eff}^{\perp}$, decreases with the increase of BFO film thickness for perpendicular geometry. Values of $H_{eff}^{\perp}$ are in a good agreement with saturation magnetization measured by VSM as shown in Fig. 4. The comparison of $H_{eff}^{\perp}$ and $4\pi M_s$ (Fig. 4) shows that the surface anisotropy $\frac{4K_{\perp}}{M_s t_F}$ term is negligible with respect to the saturation magnetization. Thus the surface anisotropy cannot explain the magnetization decrease as a function of BFO thickness (Fig. 4).

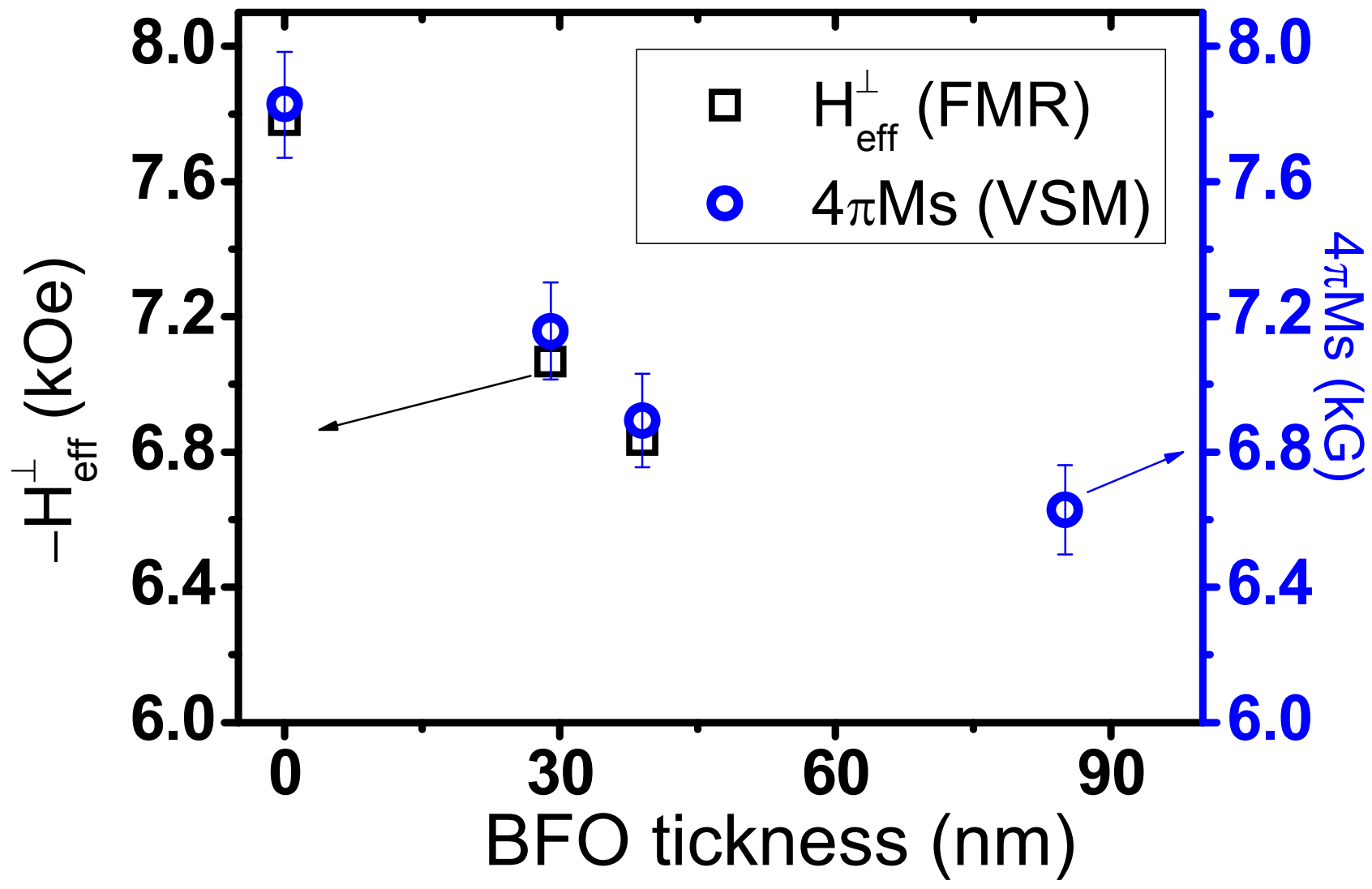

**Fig. 4** : *Evolution of the* $4\pi M_s$ *and* $H_{eff}^{\perp}$ *as a function of BFO thickness.*

Next, we determine from Figs. 3(a) and 3(b) the behaviour of gyromagnetic ratio in in-plane parallel and out-of-plane perpendicular geometries respectively. As can be seen in Fig. 5, the gyromagnetic ratio measured in perpendicular geometry $(\gamma_{\perp})$ is lower than that of parallel geometry $(\gamma_{||})$. Moreover, a monotonous decrease of $\gamma_{\perp}$ can be observed, as a function of the BFO film thickness up to $t_{BFO}$ = 39 nm. Since no FMR signal was detected beyond this thickness, we cannot extrapolate such behaviour to thicker BFO samples. The gyromagnetic ratio in parallel geometry $\gamma_{||}$ for $t_{BFO}$ = 85 nm is higher than those measured for thinner samples.

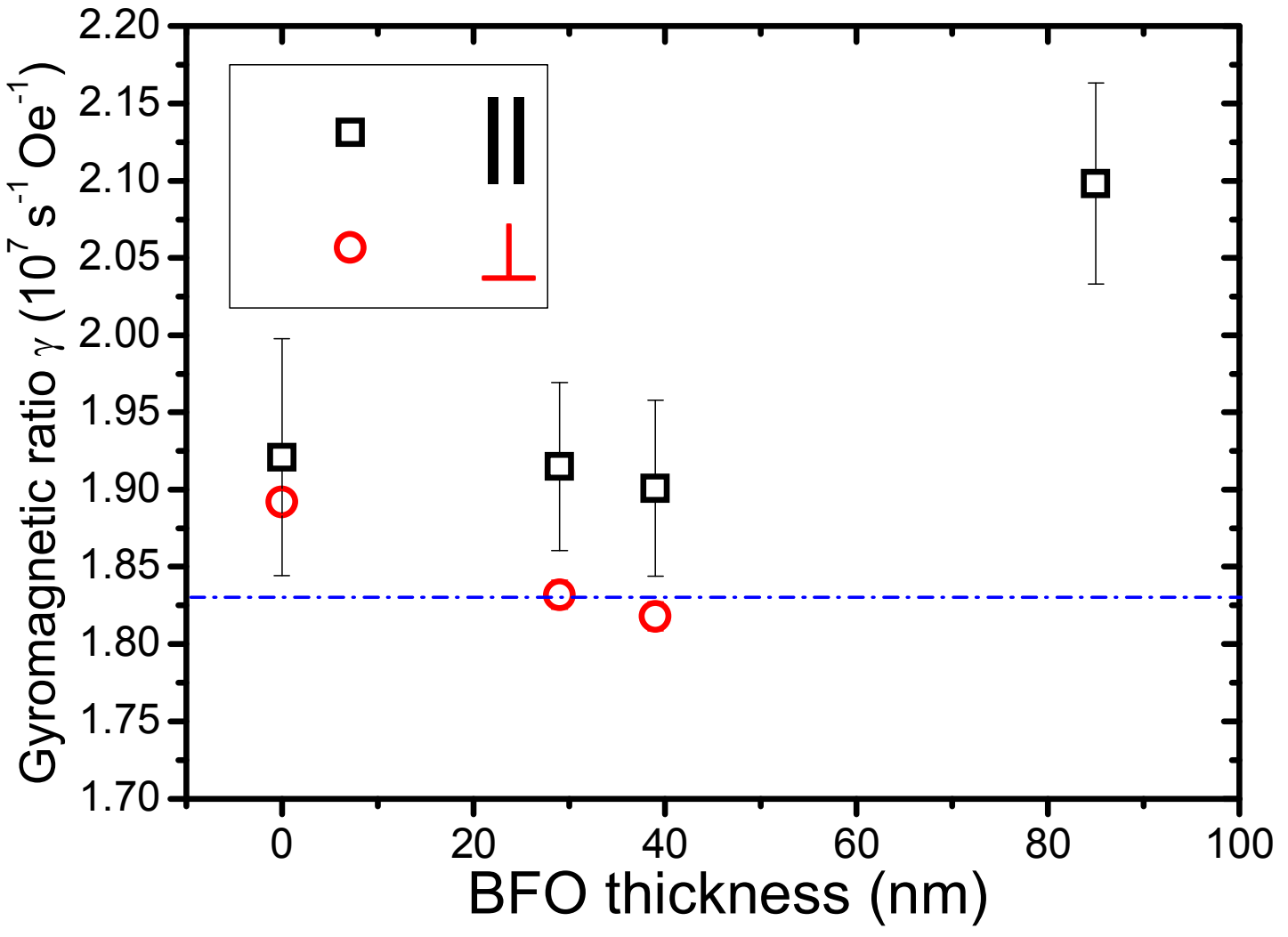

**Fig. 5** : *Behaviour of gyromagnetic ratio for in-plane parallel geometry* $\gamma_{||}$ *(along the deposition field direction) and out-of-plane perpendicular geometry* $\gamma_{\perp}$ *as a function of BFO thicknesses* $t_{BFO}$ *= 0, 29, 39, and 85 nm. The dashed line shows the gyromagnetic ratio of* $Py/Al_2O_3$ *reference sample.*

As mentioned above, the gyromagnetic ratio is a measure of the strength of spin-orbit coupling and the orbital to spin magnetic moment ratio $\mu_L/\mu_S$. Nonzero spin-orbit coupling requires a nonvanishing orbital moment and is responsible for the intrinsic magnetic anisotropy energy [39]. Bruno has shown [40] that the mean orbital magnetic moment and the magnetic anisotropy energy come from the same type of matrix element in second order perturbation theory, due to the spin-orbit coupling.

A 10-nm thick $Py/Al_2O_3$ film has a gyromagnetic ratio of $\gamma = 1.84\times10^7$ $s^{-1}Oe^{-1}$ with $4\pi M_s = 8000$ G, but when Py is sandwiched between two Pt films, $\gamma = 1.90\times10^7$ $s^{-1}Oe^{-1}$ with the same magnetization. Mizukami et al. [41] have shown that in Pt/Py/Pt structures $\gamma$ increases with decreasing Py thickness. The deviation of $\gamma$ from the free electron value depends on the strength of spin-orbit coupling. Thus, the enhancement of $\gamma$ in Pt/Py/Pt may be attributed to Pt since the latter has a large spin-orbit coupling.

In the work of Bruno [40], it was shown that the difference between planar and perpendicular gyromagnetic ratios can be understood by the presence of magnetocrystalline anisotropy [42]. In the case of thin films, the shape anisotropy arising from the dipole-dipole interaction may be larger than the intrinsic magnetocrystalline anisotropy energy and enforces the magnetization to align parallel to the thin-film plane. The magnetization, therefore, has its easy axis in the thin film plane due to the significant shape anisotropy and may not be aligned along the largest orbital moment given by the orbital magnetism, which may give rise to a difference between the perpendicular and parallel gyromagnetic ratios. In perpendicular geometry, the decrease of $\gamma$ as a function of $t_{\mathrm{BFO}}$ may be related to the vanishing of one of the Pt interfaces with Py.

In order to understand the dynamic anisotropic behaviour of the FMR field, we carried out in-plane FMR measurements as a function of azimuthal angle, shown in Fig. 6, where the deposition field direction corresponds to 0°. In Fig. 6(a) we have drawn $H_{\mathrm{res}}$ azimuthal evolution for a Pt/Py/Pt sample without BFO.

The anisotropy of Py is uniaxial with an anisotropy field of about 5 Oe [43]. In the case of Pt/Py/Pt, the azimuthal anisotropy shape [Fig. 6(a)] cannot be explained only by uniaxial anisotropy and uniform in-plane gyromagnetic ratio. The two additional lobes break the uniaxial symmetry and suggest the presence of an additional biquadratic anisotropy that may be induced by the presence of Pt which has a strong spin-orbit coupling, and/or it may also arise due to possible in-plane anisotropy of the gyromagnetic factor [20].

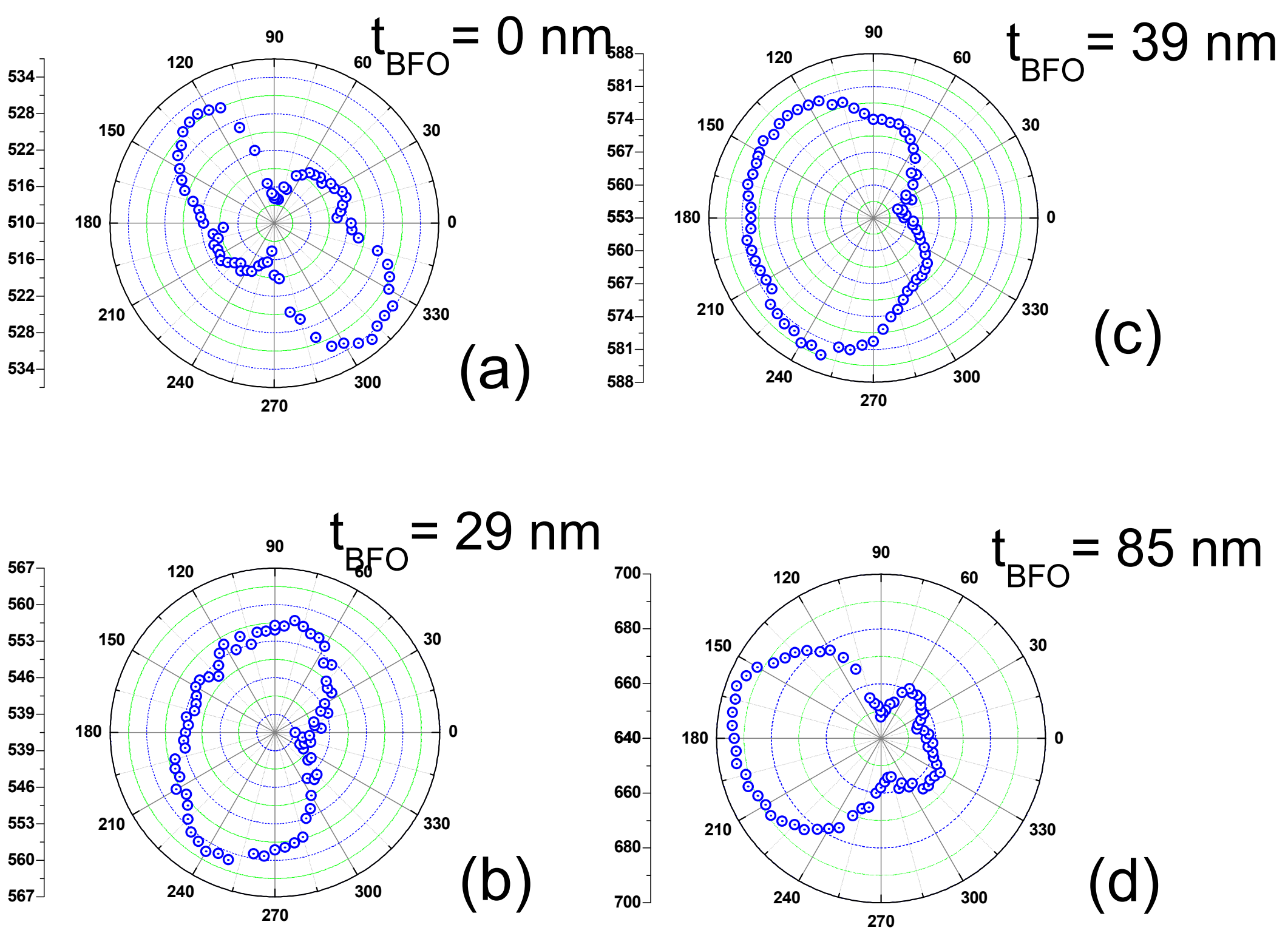

**Fig. 6** : *In plane azimuthal angular dependencies of FMR field (Oe) for different BFO thickness : (a) $t_{BFO}$= 0 nm, (b) $t_{BFO}$= 29 nm, (c) $t_{BFO}$= 39nm and (d) $t_{BFO}$= 85 nm, respectively.*

The complex azimuthal FMR field behaviour [Figs. 9(b), 9(c) and 9(d)] of our Pt/BFO/Py/Pt samples beyond the critical thickness suggests also the presence of an additional biquadratic anisotropy that may be induced by the presence of Pt but may also be induced by the presence of BFO film and the unidirectional anisotropy due to the exchange coupling [16,18]. It should be added that the presence of unidirectional, uniaxial and biquadratic anisotropies, with misaligned orientations of anisotropy axis gives rise to inexhaustible variety of azimuthal angular dependences of FMR signal as function of misalignments and anisotropy energy proportions as function of BFO thickness. In some exchange coupled bilayers, such as CoFe/NiO [29], Fe/MnPd [28] and Py/FeMn [44], the unidirectional magnetic anisotropy is accompanied by higher order magnetic anisotropies.

In these heterostructures, the anisotropy may be also influenced by surface-induced effects as well as by growth under an applied magnetic field.

## 4. Relaxation

### 4.1 Azimuthal behaviour of linewidth: polar study at f=6 GHz

The understanding of underlying physics of energy dissipation mechanisms and the control of magnetic damping is important for the development of high speed magnetic devices. The relaxation process is governed by spin-orbit interaction. Besides the FMR field, the linewidth is very sensitive to the anisotropy of the film, including surface anisotropy, and to magnetic interactions between different elements of the magnetic heterostructure. The linewidth is also sensitive to interface interactions such as exchange bias, and to the magnetic and structural inhomogeneities in the sample.

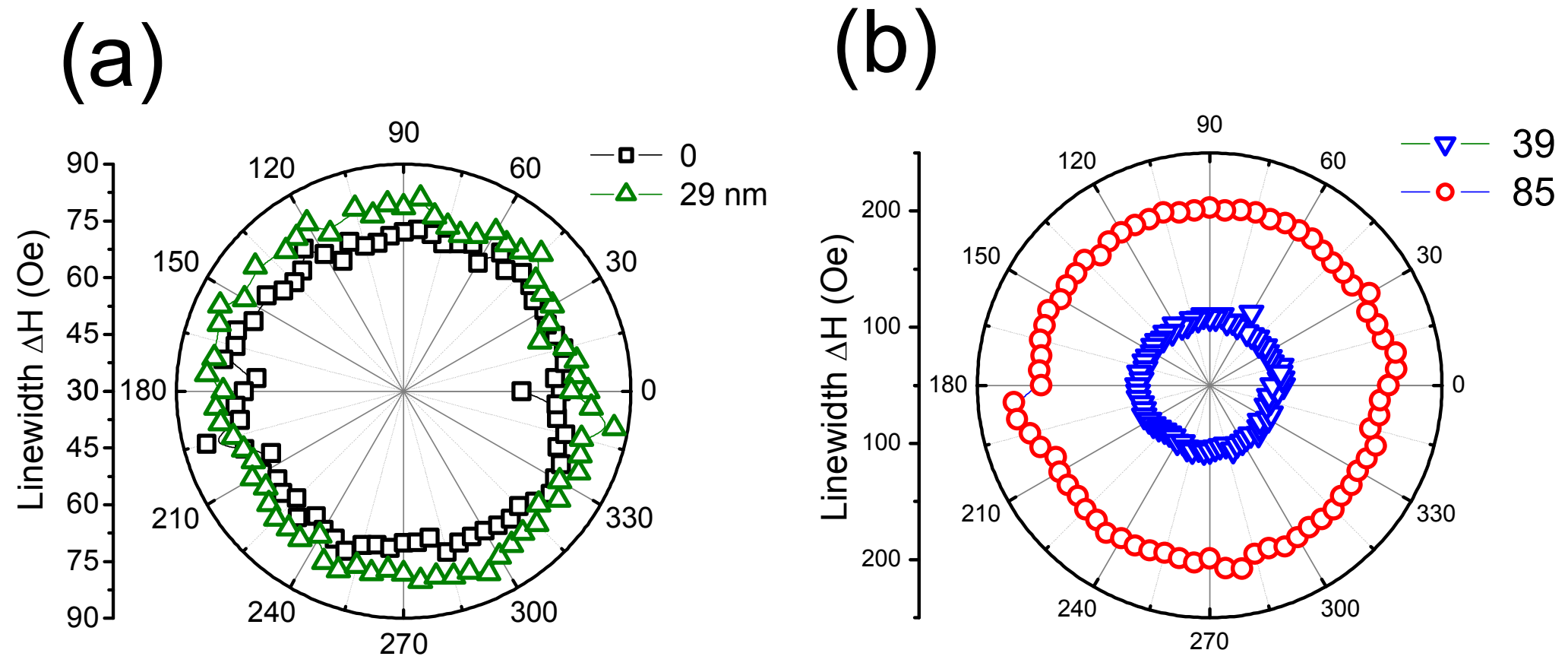

**Fig. 7** : *In-plane azimuthal angular dependence of measured FMR linewidth for different BFO thicknesses: (a)* $t_{BFO} = 0$ *and 29 nm, (b)* $t_{BFO} = 39$ *and 85 nm measured at f=6 GHz.*

For this reason, we carried out a systematic analysis of FMR linewidth dependence on BFO thickness over a wide range, including the critical thickness.

As shown in Figs. 7(a) and 7(b), the azimuthal behaviour of in-plane FMR linewidth $\Delta H$ is quasi isotropic. Note that the X-Ray diffraction patterns of our polycrystalline Pt/BFO/Py/Pt samples [16, 18] show no preferential directions and our samples are not textured.

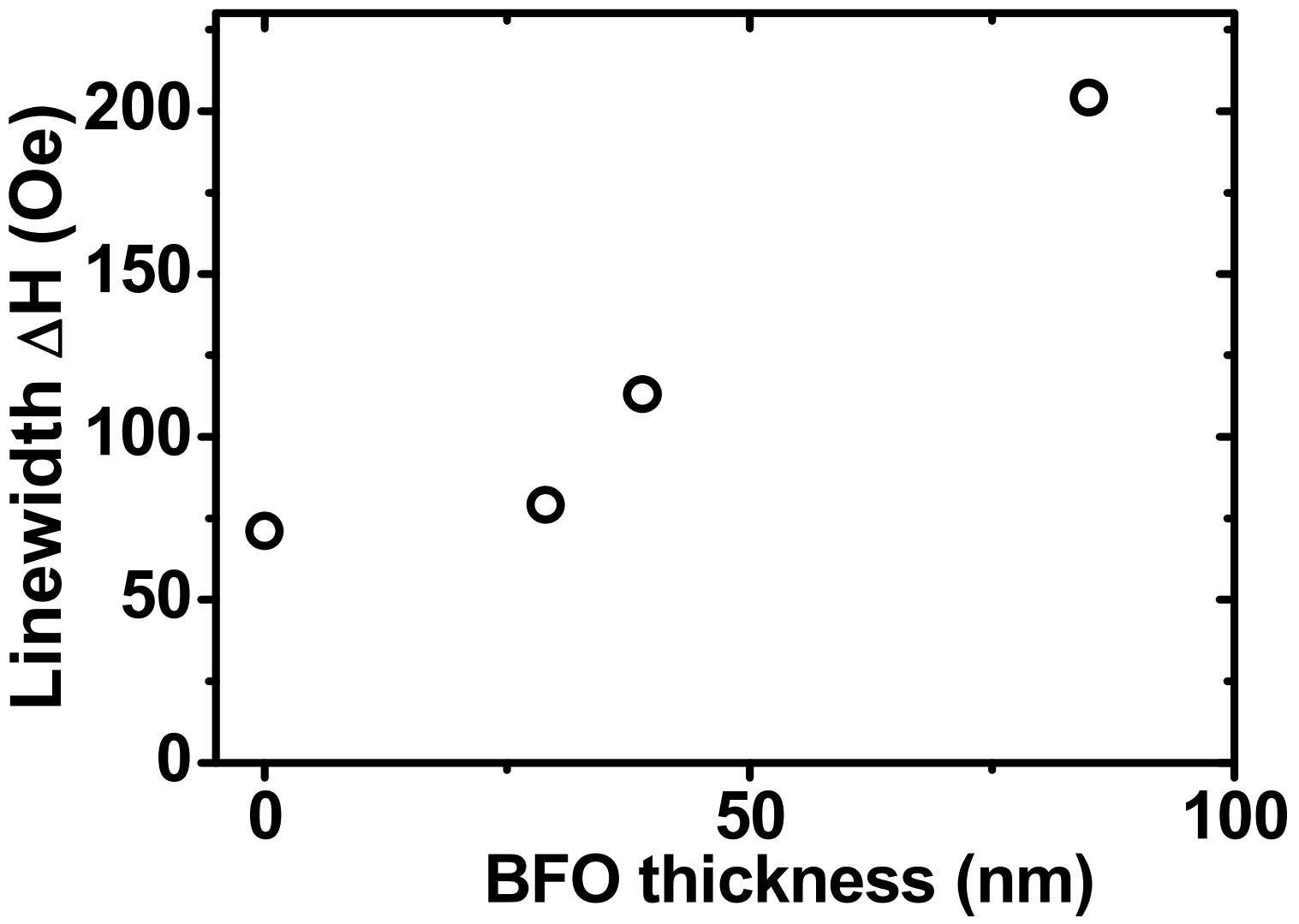

**Fig. 8** : *Evolution of in-plane FMR linewidth as a function of BFO thickness measured at f=6 GHz.*

As shown in Fig. 8, the in-plane FMR linewidth $\Delta H$ increases as a function of BFO thickness. In order to understand this behaviour we will study the $\Delta H$ dependence on the frequency in order to separate intrinsic and extrinsic components of relaxation.

## 4.2 FMR linewidth dependence on frequency in the range of 3–35 GHz

It is generally believed that the spin-orbit interaction plays a dominant role in the damping mechanism in a ferromagnet since it couples the spin to the lattice. In spite of the physical mechanism of damping, an increase in linewidth may arise from the sample inhomogeneity, so that $\Delta H(\omega)$ includes intrinsic and extrinsic contributions. The former is due mainly to the spin-orbit coupling and depends on the microwave frequency, while the latter, which is a frequency independent component, is due to the magnetic inhomogeneities.

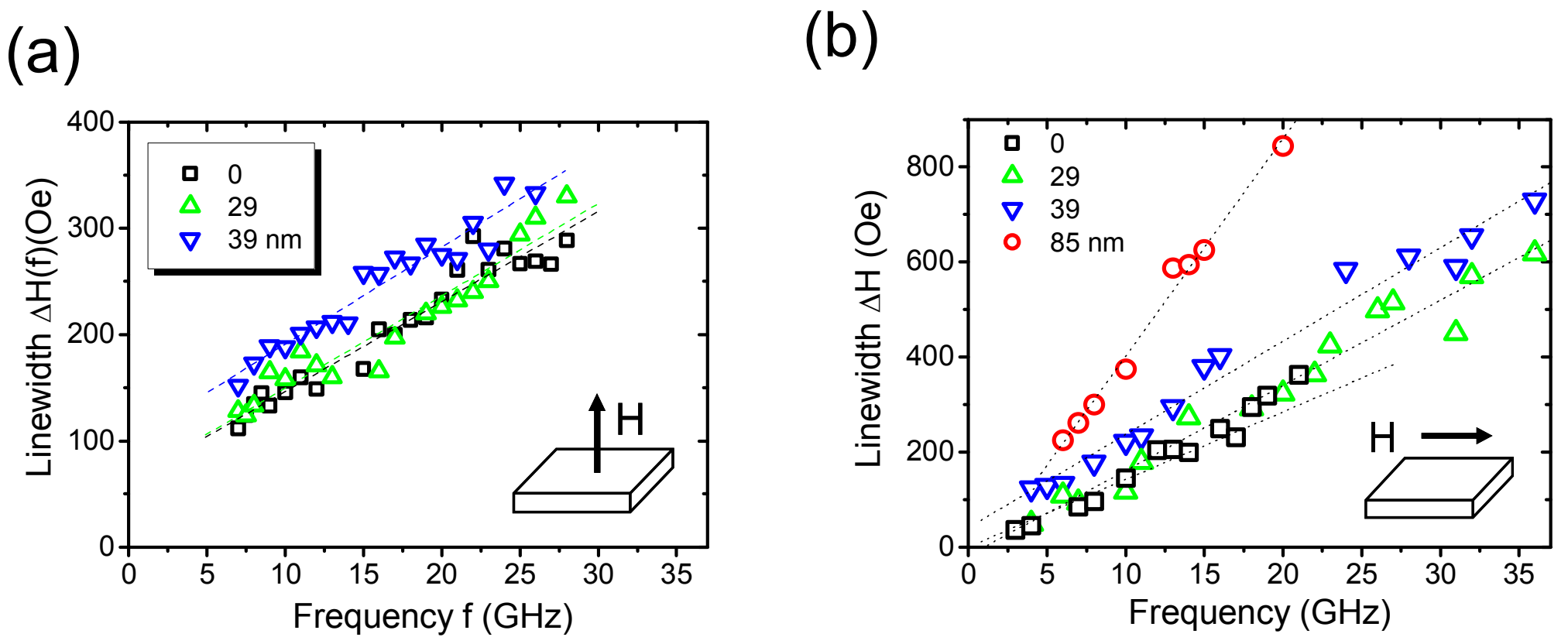

**Fig. 9** : *Evolution of the FMR linewidth in perpendicular geometry (a) and planar geometry (b) along the deposition field direction as a function of resonance frequency for different BFO thickness.*

On Fig. 9, we have measured the FMR linewidths at frequencies in the range of 3–35 GHz for out-of-plane perpendicular [Fig. 9(a)] and planar geometries in the deposition field direction [Fig. 9(b)]. The peak-to-peak FMR linewidth $\Delta H(\omega)$ of the uniform resonance mode, above several GHz, is not only proportional to the microwave angular frequency $\omega$, as expected from the Gilbert damping, but has also a zero-frequency offset $\Delta H(0)$ [45]. Thus, the frequency-dependent FMR experiments allow the extraction of different contributions to damping from a linear relation:

$$\Delta H(\omega) = \Delta H(0) + \frac{2}{\sqrt{3}} \frac{G}{\gamma^2 M_s} \omega . \qquad (3)$$

According to Suhl [45], the frequency proportional intrinsic term $\frac{2}{\sqrt{3}} \frac{G}{\gamma^2 M_s} \omega$ originates from the intrinsic damping due to the combined effect of exchange interaction and spin-orbit coupling. Thus, the FMR linewidth $\Delta H(\omega)$ has been separated into a static broadening part, $\Delta H(0)$ and a dynamic part described by the Gilbert damping rate $G$. The zero-frequency broadening $\Delta H(0)$, in the current understanding of the mechanisms that contribute to the FMR linewidth, is induced by magnetic inhomogeneities and, therefore, is extrinsic. In a nonideal sample, inhomogeneity may come from a wide variety of origins including the spread of crystallographic axis orientations in our polycrystalline sample, from lattice defects and magnetic impurities, from nonuniform stresses, from surface anisotropy with film thickness random variations, or from all other phenomena causing random variations of the internal effective magnetic fields throughout the specimen [19]. The Gilbert damping parameter $G$ is a measure of damping rate. For Ni, Co, Py and Fe films, $G$ is respectively of the order of 220, 170, 114 and 57 MHz.

On Fig. 9(a) one can see linear variation of resonance linewidth as a function of microwave frequency for different thickness of BFO. The total linewidth $\Delta H(\omega)$ in perpendicular geometry with an applied field perpendicular to the film surface as a function of frequency $f=\omega/2\pi$ follows a linear behaviour similar to in-plane parallel geometry that allows determination of the intrinsic and extrinsic contributions to the relaxation.

In in-plane parallel geometry, however, in spite of linear behaviour of the total linewidth $\Delta H(\omega)$, the term corresponding to extrinsic contribution of $\Delta H(0)$ becomes negative for Pt/Py/Pt and for samples with $t_{BFO}$ = 29 and 85 nm. For these samples, the linear approximation allowing the separation of $\Delta H(\omega)$ into extrinsic and intrinsic terms becomes inappropriate. It should be noted that for microwave frequency below $f$=4 GHz, the absorption of uniform mode is close to the absorption peak of the unsaturated mode [36-37]. The energy dissipation of the uniform mode either by two-magnon scattering or by other dissipation channels may invalidate the linear dependence of linewidth on the frequency expressed by Eq. (3) [46]. It should also be noted from Fig. 9(a) that the FMR signal vanishes for $t_{BFO}$=85 nm in out-of-plane perpendicular geometry.

From Fig. 9, we have deduced the intrinsic dimensionless Landau-Lifshitz damping parameter for perpendicular $\alpha_{\perp}$ and parallel geometries $\alpha_{||}$ as well as $\Delta H_{\perp}(0)$ and $\Delta H_{||}(0)$ by using the relation $G = \alpha\gamma M_s$. It can be seen on Fig. 9, for $t_{BFO}$ = 39 nm, $\alpha_{||}$ =0.049 ($\Delta H_{||}(0)$ =40.5 Oe), which is rather high compared to its usual value (without a Pt environment) which is $\alpha_{||}$ = 0.01 [47]. Mizukami et al.[48] have shown a strong dependence of FMR linewidth in nonmagnetic/Py/nonmagnetic heterostructures on the origin of nonmagnetic metal sandwiching Py film. In particular, Pt and Pd present the highest value for Gilbert damping parameter $G$, due to the large spin-orbit coupling at interfaces. A mechanism that can explain the enhanced damping in Py/Pt is the spin-pumping at the interface of Pt and Py [26]. The magnetization precession in Py drives a spin current into the adjacent Pt film where it relaxes rapidly since Pt is a strong spin scatterer (spin-diffusion length of the Pt is $\lambda_{Pt}$ = 10 nm at room temperature [49]). In our samples Py is sandwiched between BFO film and Pt buffer film. The high damping of the sample with $t_{BFO}$ = 39 nm is due to both exchange coupling enhancement at the BFO interface and to a spin-pumping mechanism at the Pt interface [50-51].

The intrinsic damping parameter $\alpha_{\perp}$ in out-of-plane perpendicular geometry is usually lower than that of the in-plane parallel geometry [52], for $t_{BFO}$ = 39 nm $\alpha_{\perp}$ = 0.018 ($\alpha_{||}$ = 0.049). The damping dependence on the applied static magnetic field direction leads the damping parameter difference in both geometries to be related to the two-magnon scattering processes. The uniform mode ($k$ = 0) may transfer its energy efficiently to spin waves with ($k$ ≠ 0) if there are degenerate states available for scattering. It is well known that, for perpendicular geometry, the FMR frequency lies near the bottom of the spin-wave manifold and there are few degenerate states. In the out-of-plane parallel geometry, the spin-wave manifold shifts down in frequency in such a way that the FMR frequency is at the top of the band and there are more degenerate states available [52].

The difference of the intrinsic damping parameter $\alpha$ of about a factor of two (for the sample $t_{BFO}$ = 39 nm) between out-of-plane perpendicular and in-plane geometries cannot be explained only by spin-wave damping differences. Several authors have observed important damping differences in these two geometries in Pt/Py systems with planar anisotropy [53-55], which was attributed to the spin-pumping dependence on the applied field direction. In our system, the factor of two damping difference may be attributed to the spin-pumping effect since, without the Pt cap and buffer layers, the difference is an order of magnitude smaller than observed.

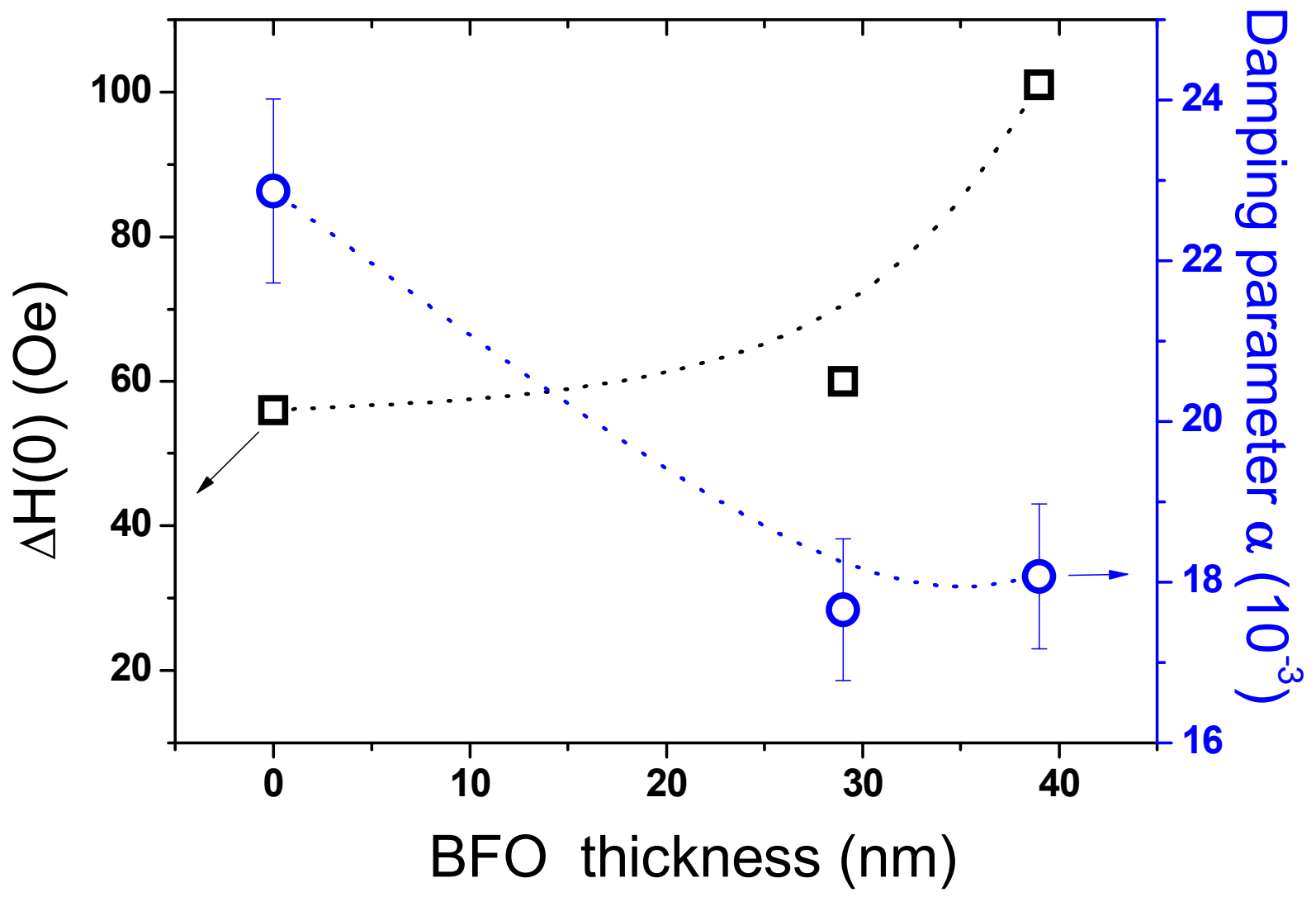

**Fig. 10** : *Evolution of the zero-frequency broadening ΔH(0) and the intrinsic damping parameter* $\alpha$ *as a function of BFO thickness for out-of-plane perpendicular geometry.*

When insulating BFO film is inserted between Py and the Pt buffer layer, the spin-pumping increase of intrinsic damping parameter occurs only on the interface of Py with the Pt cap layer. Further increase of BFO thickness up to the critical thickness attenuates the spin-pumping induced damping as shown in Fig. 10. However, beyond the critical BFO thickness, the intrinsic damping parameter increases due to the exchange coupling as observed in different exchange biased systems.

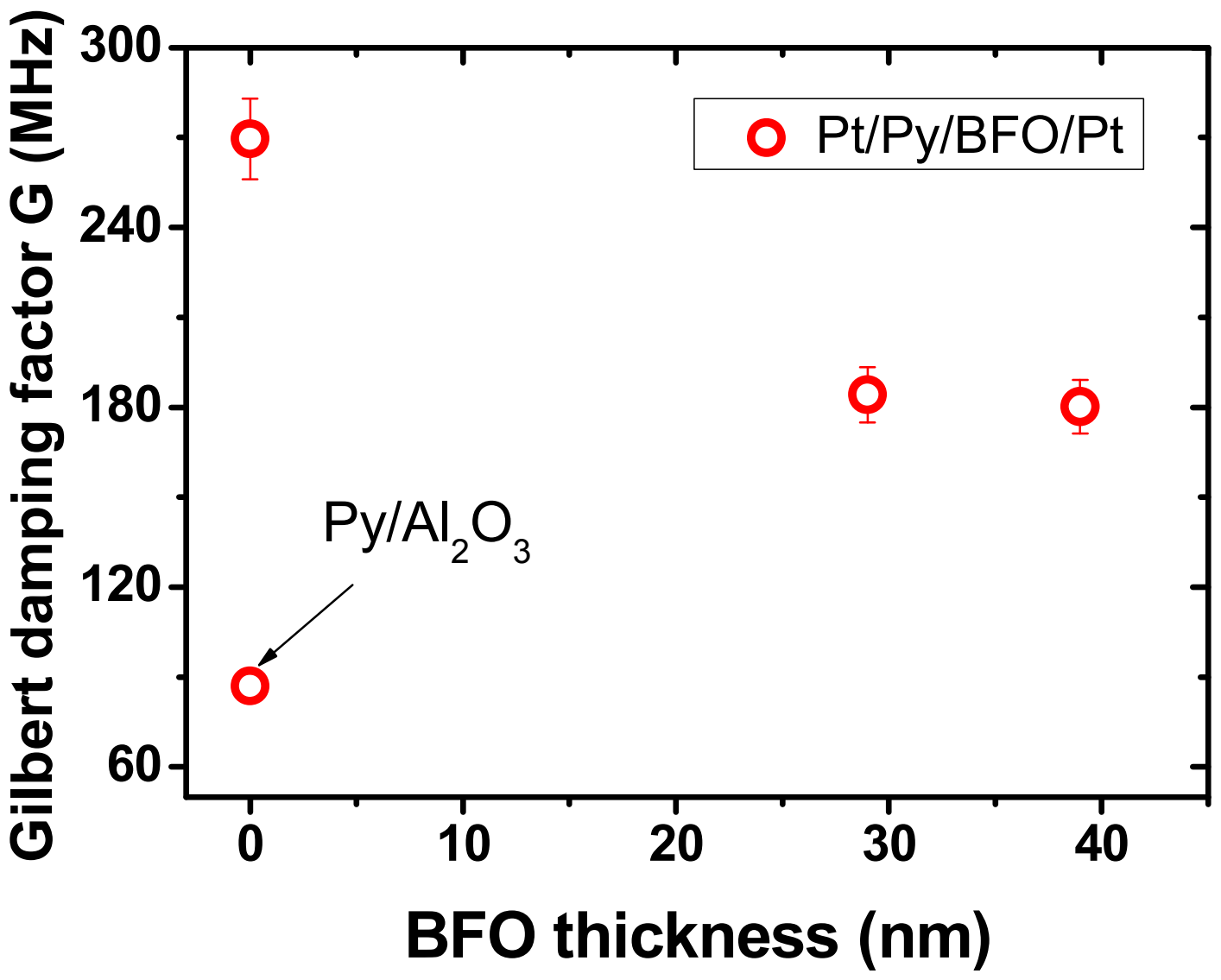

**Fig. 11** : *Evolution of Gilbert damping factor G, as a function of BFO thickness. The value of G of the Py/$Al_2O_3$ reference sample is added to the graph.*

In our system, the saturation magnetization $4\pi M_s$, the gyromagnetic ratio, the Gilbert damping rate $G$ and the dimensionless intrinsic damping parameter $\alpha$ depend on the BFO thickness. They are also sensitive to the strength of exchange coupling, the roughness, the magnetic anisotropy and the chemical composition of the buffer and cap layers. In some theories, $\alpha$ is predicted to be inversely proportional to $M_s$. For the comparison of results, it is convenient to use the Gilbert damping rate $G$ since it allows normalization of results for different $M_s$ and $\gamma$ values.

The Gilbert damping rate $G$ is more instructive than the intrinsic damping parameter $\alpha$. In Fig. 11, we show Gilbert damping $G$ evolution with BFO thickness in the Pt/Py/BFO/Pt/ system, compared to the value of a reference Py/$Al_2O_3$ sample. In order to determine the effect of the Pt cap layer on the damping properties of Py, Py/$Al_2O_3$ has been grown and compared with Py/Pt and Py/BFO/Pt heterostructures.

Here, one can see a strong increase of $G$ in the presence of Pt on interfaces with Py. $G$ decreases about 30% when a BFO film is inserted between Py and Pt on one of the interfaces. Further increase of the BFO insert does not substantially change the value of $G$. These results clearly show that the magnetic relaxation in our system is dominated by the spin-pumping mechanism due to the presence of Pt via strong spin-orbit interactions on the interface with Py. However, for thicker BFO inserts, an additional damping mechanism appears due to the exchange bias, beyond the critical thickness, above which a hysteresis curve shift is observed.

## Conclusion

By performing ferromagnetic resonance measurements, we have analysed dynamic properties of polycrystalline exchange-biased Pt/BFO/Py/Pt systems. Practically identical saturation magnetization and effective field were found, which implies negligible surface anisotropy and dipolar fields.

In-plane azimuthal measurements of FMR field revealed unusual behaviour with the appearance of two additional lobes suggesting the presence of an additional biquadratic anisotropy or/and an in-plane anisotropy of the gyromagnetic ratio.

The gyromagnetic ratio measured in out-of-plane perpendicular geometry was observed to be lower than that in parallel geometry. The monotonous decrease of gyromagnetic ratio in out-of-plane perpendicular geometry as a function of the BFO film thickness was related to the vanishing of one of the Pt interfaces with Py. The enhancement of $\gamma$ in Pt/Py/Pt was attributed to the Pt because of its large spin-orbit coupling. The behaviour of the gyromagnetic ratio in parallel geometry was interpreted by contributions of Pt substitution on one interface by a BFO film and also by the appearance of exchange coupling.

The total FMR linewidth of uniform mode has an almost isotropic azimuthal shape and increases as a function of BFO thickness. The separation of FMR total linewidth into a static broadening part, and a dynamic part described by Gilbert damping rate *G* allowed the distinction of physical mechanisms contributing to the FMR linewidth by magnetic inhomogeneities and by intrinsic dissipation channels. In our system, the difference of a factor of two in damping parameter between planar and perpendicular geometries was attributed to the spin-pumping effect since, without a Pt cap and a buffer layer, the difference is an order of magnitude smaller than observed. These results clearly show that the magnetic relaxation in our system is dominated by the spin-pumping mechanism due to the presence of Pt via strong spin-orbit interactions on the interface with Py. An additional damping mechanism for higher BFO thickness appears due to the exchange bias for thicknesses beyond the critical thickness.

**Acknowledgments**

The authors thank N. Vukadinovic for valuable discussions and R. Tweed for his help in the preparation of the manuscript.